\documentclass[twocolumn,prb,aps,showpacs,floatfix,amsmath,superscriptaddress]{revtex4}

\usepackage{graphicx}% Include figure files

\begin{document}

\title{Optimizing transport efficiency on scale-free networks through assortative
 or dissortative topology}

\author{Yu-hua Xue}
\affiliation{College of Physical Science and Technology, Yangzhou University, Yangzhou 225002, P. R. China  }

\author{Jian Wang}
\email{wangjian@yzu.edu.cn}
\affiliation{College of Physical Science and Technology, Yangzhou University, Yangzhou 225002, P. R. China  }
\affiliation{Department of Physics, Centre for Nonlinear Studies and The Beijing-Hong Kong-Singapore Joint Centre for Nonlinear and Complex Systems (Hong Kong), Hong Kong Baptist University, Kowloon Tong, Hong Kong, China }

\author{Liang Li}
\affiliation{College of Physical Science and Technology, Yangzhou University, Yangzhou 225002, P. R. China  }

\author{Daren He}
\affiliation{College of Physical Science and Technology, Yangzhou University, Yangzhou 225002, P. R. China }

\author{Bambi Hu}
\affiliation{Department of Physics, Centre for Nonlinear Studies and The Beijing-Hong Kong-Singapore Joint Centre for Nonlinear and Complex Systems (Hong Kong), Hong Kong Baptist University, Kowloon Tong, Hong Kong, China }
\affiliation{ Department of Physics, University of Houston, Houston, Texas 77204-5005, USA }

\date{22 Feb. 2010}

\begin{abstract}
 We find that transport on scale-free random networks depends strongly on degree-correlated network topologies whereas transport on Erd$\ddot{o}$s-R$\acute{e}$nyi networks is insensitive to the degree correlation. An approach for the tuning of scale-free network transport efficiency through  assortative or dissortative topology is proposed. We elucidate that the unique transport behavior for  scale-free networks results from the heterogeneous distribution of degrees.
\end{abstract}

%\keywords{ transport efficiency, scale-free networks, degree-degree correlation }
\maketitle

\emph{Introduction.} Understanding transport process on networks\cite{basreview} is  a central problem
in many fields,\cite{basreview,rev1} ranging from social networks to natural or
technique networks. For example, epidemic spreading\cite{epdemicl,epdemic2}  and
transportation\cite{hwoongprl} are typical phenomena of transport related to social
networks, while the World-Wide Web, the Internet, the biological
networks and the new network-based materials are technical or
natural networks.\cite{rev1,epdemicl,epdemic2,hwoongprl,electricnet} Usually, we can classify transport on networks into two categories according to whether the network flow conservation is
observed for each node of  networks: the contamination process\cite{epdemicl,epdemic2}
and the network flow problem.\cite{hwoongprl,electricnet,ahujabook,lopez1,carmi1,guangliang1}

The network topology has a profound implication for network transport.\cite{basreview,rev1,epdemicl,epdemic2,lopez1,carmi1,electricnet,ahujabook,guangliang1,hwoongprl} One of the important topological features in the network structure is the tendency of vertices with a given degree to be
connected to other vertices with similar degree(assortativity)  or dissimilar degree(dissortativity).\cite{newman1} Many real networks exhibit this degree correlation among their nodes.  Such correlation plays an important role in transport process on networks.  For the epidemic spreading, lack of an epidemic threshold has been verified in assortative networks.\cite{epdemic2}  However, in  the network flow problem, the role of such topological vertex correlation is still  unclear. Understanding the role of the degree correlation on the network flow process is important not only to biological networks\cite{basreview,rev1} and traditional transportation networks\cite{hwoongprl},  but also to the design of new network-based materials.\cite{electricnet}  Thus, some interesting questions are:  the degree correlation improve or deteriorate network transport? Can network transport efficiency be optimized through assortative
 or dissortative topology?

 In this report, we first present the results of transport on scale-free random networks and Erd$\ddot{o}$s-R$\acute{e}$nyi networks with different degree-correlated topologies. Then, a further comparison with two empirical networks is made.

\emph{Transport on networks.}  Let $G=(N,A)$ be a network defined by a set N of nodes and a set A
of edges.  Each edge $(i,j)$ connected from  node $i$ to node $j$  has
an associated cost function $c_{ij}$, which denotes the cost per unit flow on
that edge. With the quadratic cost function, we can solve the maximum flow problem through Kirchhof's equations.\cite{ahujabook,kirchhoff} This quadratic restriction on the cost function captures the essential properties of many important physical networks.\cite{hwoongprl,ahujabook}  Then these equations are solved using the itpack method\cite{sparsematrix} parallelly.   To characterize the transport capability of a whole network,\cite{Latora1} we average the network conductance $G_{st}$ between each pair of nodes as
\begin{equation}
 \label{aveg}
 \big<G\big> = \frac{1}{N(N-1)}\sum_{s,t\in N; s\neq t}{G_{st}},
\end{equation}
where $s$ and $t$ run from $1$ to $N$. Larger average conductance $\big<G\big>$ signifies a better transport capability of networks.  From a statistical perspective,  a probability density function\cite{lopez1}(pdf) $\phi (G)$ can be defined through $\phi (G)dG$,  which denotes the probability that two nodes have conductance between $G$ and $G+dG$. A cumulative distribution function(cdf) $F(G)$ can be described by $F(G)= \int_{G}^{\infty}{\phi(G')dG'}$.

 We use the uncorrelated configuration model\cite{ucfmethod} to generate the uncorrelated scale-free networks with the degree distribution$P(k) \sim k^{-\gamma}$,  where $k$ is the number of links attached to the node. In fact, scale-free networks generated from the uncorrelated configuration model belong to the scale-free random networks. Uncorrelated Erd$\ddot{o}$s-R$\acute{e}$nyi networks are constructed with the standard random method.\cite{randomnets} We further employ the algorithm of reshuffling links\cite{reshuffling1,reshuffling2}  to transform the existing uncorrelated network to correlated networks. During the following calculation, we utilize the Newman factor\cite{newman1} and the mean nearest neighbor function\cite{newman1}$K_{nn}(k)$ to measure the degree-degree correlation property of networks.

\emph{Scale-free random networks.} The scale-free networks with size $N=3000$ are constructed using the aforementioned method. To simplify the calculations, we set the value one to the weight of each edge in the networks, $v_s=1$ and $v_t=0$ for the source and sink nodes, respectively.
\begin{figure}[t]
\includegraphics[width=0.90\columnwidth]{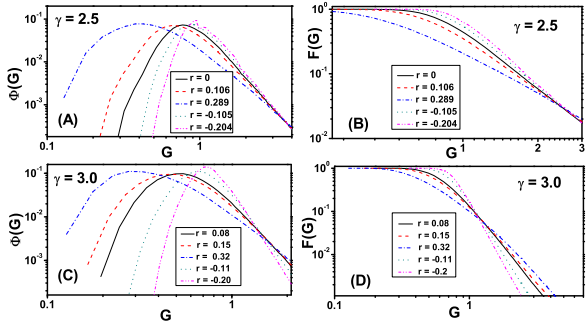}
\caption{\label{fig:distribution} (Color online). The pdf $\phi(G)$ and the cdf $F(G)$  vs $G$ for different degree-correlated scale-free networks with $\gamma=2.5$ and $\gamma=3.0$. The values of Newman factors $r$ are illustrated in each figure.  }
\end{figure} The pdf $\phi (G)$ and cdf $F(G)$ vs $G$ for scale-free networks with different degree correlations are illustrated in Fig.~\ref{fig:distribution}.
\begin{figure}[b]
\includegraphics[width=0.90\columnwidth]{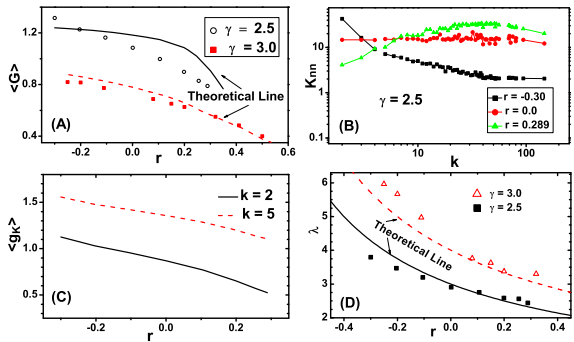}
\caption{\label{fig:phonondisp1} (Color online). $\bf(A)$ Average network conductance $\big<G\big>$ as a function of the Newman factor $r$. Solid line and dash lines are theoretical results calculated from Eq.~(\ref{avecond}). $\bf(B)$ The mean nearest neighbor function $K_{nn}(k)$ vs the degree $k$ with different degree correlations. $\bf(C)$ Degree-averaged network conductance $g_k$ changes with $r$. $\bf(D)$ The numerically fitted exponents $\lambda$ for the power-law tail against r.  Solid and dash lines are theoretical predictions. }
\end{figure}

It can be seen from Fig.~\ref{fig:distribution}(A) and Fig.~\ref{fig:distribution}(C) that the peaks for the pdf $\phi (G)$ shift to higher  $G$ when the topological structures vary from assortativity to dissortativity. The change of average network conductance $\big<G\big>$ against the degree correlation coefficient(Newman factor) $r$ is plot in Fig.~\ref{fig:phonondisp1}(A). It can be found that $\big<G\big>$ is significantly reduced  when the network topology is assortative.  In contrast, the network conductance has been remarkably improved for the dissortative topology.  To quantify such variation of average network conductance with the degree-mixing properties, we introduce the \textit{metrics of network transport efficiency} as
\begin{equation}
\label{efficiency}
\eta=\frac{\big<G(r)\big>-\big<G(r=0)\big>}{\big<G(r=0)\big>},
\end{equation}
where $\big<G(r)\big>$ is average conductance for the network with the Newman factor $r$ and $\big<G(r=0)\big>$ is average network conductance for the uncorrelated networks. For $\gamma=2.5$, we can find that the network transport efficiency $\eta$ decreases by $27\%$ at $r=0.289$,  in contrast with an increase of  $21\%$  at $r=-0.30$ as shown in Fig.~\ref{fig:phonondisp1}(A).  The scale-free network with $\gamma=3.0$ shows a similar behavior. Thus we can propose that the transport efficiency on scale-free networks be controlled by tuning the dissortative/assortative network topology.

The reason why  transport efficiency on scale-free networks  depends strongly on the degree correlation lies in the fact that \textit{the network conductance between low-degree nodes can be effectively changed through tuning the degree correlation topology}.  To demonstrate this, in Fig.~\ref{fig:phonondisp1}(C) we plot the change of the degree-averaged network conductance against the Newman factor $r$ for the small-degree nodes. Here the degree-averaged network conductance $\big<g_k\big>$ means that the network conductance is averaged over the pairs of nodes, where either or both nodes have the degree $k$. It can be seen that $\big<g_k\big>$ for both $k=2$ and $k=5$ show an evident decrease for the assortative topology, in contrast with a significant increase for the dissortative topology.

The straightforward understanding for such a change of network conductance for the low-degree nodes with the degree correlation topology is as follows. For the dissortative topology, nodes with small degrees tend to be connected to other high-degree nodes such that more parallel branches are possible for the in/out-flows to/from the low-degree nodes. Therefore, the network conductance for the low-degree nodes under dissortativity  is enhanced due to the presence of the neighboring high-degree nodes.  In contrast, for the assortative topology, nodes with small degrees appear to be connected with the similarly low-degree nodes, which results in a reduced number of possible parallel flow branches. Thus, the network conductance for the low-degree nodes with the assortative topology decrease correspondingly. To verify this, the mean nearest neighbor functions $K_{nn}(k)$ is plotted in Fig.~\ref{fig:phonondisp1}(B) for the uncorrelated and correlated scale-free networks.

 To document the above analysis, we  further carry out a heuristical derivation of network conductance in the picture of transport backbone using a simplified branching process.\cite{backbone} If we ignore the loops on networks, the conductance between node $A$ and node $B$ can be modeled\cite{backbone} by a transport backbone with the average branching factor $\beta(k)=K_{nn}(k)-1$.  In fact, the neglect of loops is reasonable when the networks considered are sparsely connected.  So it is straightforward that the conductance $g_a$ between node $A$ and the infinite distance can be derived from the recursion relations\cite{backbone,guangliang1} as $g_a =  k(1-{1}/{\beta}).$ For scale-free networks, we only need to concentrate on the nodes with small degrees because most nodes are distributed in the range of low-degrees. So we assume $\beta(k)\sim \alpha k^{r}-1$, where $\alpha$ is a constant, depending on the degree-mixing property of networks and $r$ is the Newman factor. It can be seen from Fig.~\ref{fig:phonondisp1}(B) that this relation holds for small values of degrees. Thus, the conductance $G_{ab}$ between node $A$ and node $B$ can be approximately written as $G_{ab}=g(k)/2$, provided that the values of $k_a$ and $k_b$ are small. The average conductance $\big<G\big>$ on the network therefore can be written as
\begin{equation}
\label{avecond}
\big<G\big>\simeq \frac{1}{2}\int_{k_{min}}^{k_c} g(k) \cdot P(k)dk,
\end{equation}
where $k_c$ is the cutoff value for small degrees and $P(k) \sim \;k^{-\gamma}$. The calculated average network conductance by Eq.~(\ref{avecond}) is plotted in Fig.\ref{fig:phonondisp1}(A). We can find that the calculated tendencies are qualitatively consistent with the numerical simulations.  The discrepancy between numerical and theoretical results may come from loops and  specific topology structures, which have been neglected in the branching process.

When we average the scale-free network conductance between each pair of nodes, the contributions from the low-degree nodes are dominant by virtue of large numbers of low-degree nodes. Therefore, network transport efficiency on scale-free networks changes correspondingly with the degree correlation. Similar reason accounts for the peak shift for the pdf $\phi (G)$  as shown in Fig.~\ref{fig:distribution}.

In addition, a power-law tail distribution of network conductance can be observed in both pdf $\phi(G)$ and cdf $F(G)$ as illustrated in Fig.~\ref{fig:distribution}.  Such a power-law tail has been reported in Ref.~\onlinecite{lopez1} for the uncorrelated scale-free networks.  Here we find the scaling exponent of the power-law tail is related to the degree correlation.  In Fig.~\ref{fig:phonondisp1}(D), we fit the cumulative power-law distribution as $F(G)\sim G^{-\lambda}$. The fitted power-law exponent $\lambda$ clearly shows a tendency to decrease with the Newman factor $r$.

We can understand this variation of $\lambda$ with the degree correlation through an analytical derivation.  Owing to the fact that the distribution of conductance $G_{ab}$ is characterized by the distribution of nodes with small degrees,  the conditional probability\cite{lopez1} $Pr(k_a\!<\!k_b)$ with the constraint $k_a<k_b$ can be described as $Pr(k_a<k_b)dk_{a}\propto k_a^{-2\gamma+1}dk_{a}$. Here, we further make an explicit assumption for $g_a$ associated with $r$  as $g_a\sim k_a^{1+r}$, i.e. $G_{ab}\sim k_a^{1+r}$ for small-degree nodes. Such assumption agrees with the the maximum-flow problem which states that the average maximum-flow between a pair of nodes is proportional to the smaller degree $k$.\cite{dslee}  We thus obtain distributions of the cdf $F(G)\sim G^{-(2\gamma-2)/(1+r)}$. It can be seen from Fig.~\ref{fig:phonondisp1}(D) that the predictions of $\lambda_{theory}=(2\gamma-2)/(1+r)$ shows the similar tendency as the fitted results.

\begin{figure}[h]
\includegraphics[width=0.90\columnwidth]{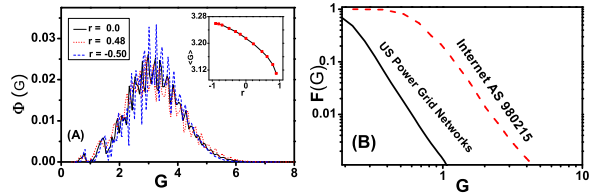}
\caption{\label{fig:randomnetworks} (Color online). $\bf{(A)}$ The pdf $\phi(G)$ for Erd$\ddot{o}$s-R$\acute{e}$nyi networks with different degree correlations. The inset shows the change of average network conductance as a function of Newman factor $r$. $\bf{(B)}$ The cdf $F(G)$ for two empirical networks: the internet structure\cite{asnetwork} at the autonomous system level (AS980215) and the U.S. power grid networks\cite{pgnetwork}. }
\end{figure}

\emph{Erd$\ddot{o}$s-R$\acute{e}$nyi random networks.}  Network conductance for Erd$\ddot{o}$s-R$\acute{e}$nyi networks with size of $N=3000$, and average degree $\bar{k}=8$ is plotted in Fig.\ref{fig:randomnetworks}(A).
In contrast with scale-free networks, pdf $\phi(G)$ and average network conductance $\big<G\big>$ for Erd$\ddot{o}$s-R$\acute{e}$nyi networks show no significant dependence on degree correlation topologies. Although the tuning of degree correlation can induce some changes of  network conductance for small-degrees nodes, but most nodes are distributed around the average degree $\bar{k}$ due to the Poisson degree distribution. Therefore, the contributions from the low-degree nodes are trivial such that the tuning of degree correlation does not significantly alter the average network conductance.

\emph{Empirical networks} We calculate network conductance in two empirical networks: the Autonomous Systems(AS)level of the Internet and the U.S. Power Grid(PG) networks. The degree distributions of the two networks show a scale-free characteristics, with $\gamma=2.21\pm0.08$ for the AS network and $\gamma=3.7\pm0.2$ for the power grid networks. The Newman factors $r$ are $r_{AS}=-0.2149$ and  $r_{PG}=-0.03$, respectively.  In Fig.~\ref{fig:randomnetworks}(B), the fitted exponents $\lambda$ of $F(G)\sim G^{-\lambda}$ are $\lambda_{AS}=2.46\pm0.05$ and $\lambda_{PG}=4.74\pm0.08$, respectively.  We theoretically calculate the scaling exponent $\lambda$ as $\lambda_{AS}^{theory}=3.08$ and $\lambda_{PG}^{theory}=5.24$. The results for empirical networks are  qualitatively consistent with  that of the scale-free random network model. The large discrepancy between the real networks and the theoretical predictions  may come from specific network topologies and loops that have been neglected in the theoretical formula.

In summary, we have found that transport on scale-free random networks vary significantly with the degree correlations. An approach for the tuning of transport efficiency on scale-free networks through dissortative or assortative topology is proposed. We elucidate that the unique transport behavior for scale-free networks results from the heterogeneity of the degree distributions. We believe that our results provide some insights into the design of network transport.

J. Wang would like to thank Eduardo L$\mbox{\'o}$pez for helpful discussions on Ref.~\onlinecite{lopez1}.  The authors acknowledge the support from  National Natural Science Foundation of China (NSFC) under the grant 10705023 and 10635040, as well as from Jiangsu Natural Science Foundation under the grant BK2009180.

\end{document}